\documentclass[american]{paper}
\usepackage{color}
\usepackage{amsmath}
\usepackage{setspace}
\usepackage[numbers]{natbib}
\usepackage[unicode=true]
 {hyperref}

\makeatletter
\newenvironment{lyxcode}
	{\par\begin{list}{}{
		\setlength{\rightmargin}{\leftmargin}
		\setlength{\listparindent}{0pt}
		\raggedright
		\setlength{\itemsep}{0pt}
		\setlength{\parsep}{0pt}
		\normalfont\ttfamily}%
	 \item[]}
	{\end{list}}

\makeatother

\begin{document}
\title{Using \emph{Maple} and \emph{GRTensorIII}\, in relativistic spherical
models}
\author{Medina V.\thanks{Departamento de Fisica - Facultad de Ingenieria - Universidad de Carabobo.
Valencia, Venezuela: \protect\href{mailto:vmedina@uc.edu.ve}{vmedina@uc.edu.ve}} }
\maketitle
\begin{abstract}
This article presents some aspects and experience in the use of algebraic
manipulation software applied to general relativity. Some years ago
certain results were reported using computer algebra platforms, but
the growing popularity of graphical platforms such as Maple allows
us to approach the problem of the simplifications of many expressions
from another point of view. Some simple algebraic programming procedures
are presented (in Maple with the GRTensorIII package) to obtain and
study material distributions with spherical symmetry and to search
for exact solutions of the Einstein field equations. The purpose is
to show how useful a computer algebra system can be. All calculations
were performed using the GRTensorIII computer algebra package, which
runs on Maple 2017, along with several Maple routines that we have
used specifically for the simplification of many of the algebraic
expressions that are very common in this type of problem.
\end{abstract}
\bigskip{}


\pagebreak{}

\section{Introduction }

Computer algebra systems (CAS) have a wide variety of applications
in fields that require time-consuming, difficult-to-perform, and error-prone
calculations when done manually. They are used more and more frequently
and especially when it is necessary to complete a calculation in pages
after pages for many hours or perhaps several days. As early as the
late 1950s and early 1960s, various programs appeared \citep{1953MsC..MIT,1953TempleUniv..MsC,1960CACM..3.4..184M,1964ACM64..19..19B}
aimed at demonstrate that in the scientific field it is possible to
go beyond the purely numerical area and use them to carry out a symbolic
calculation. For reasons such as these, various members of the computer
algebra family of languages were created, and in fact many were created
to perform specific calculations of great complexity in fields such
as electronic optics \citep{2017Else..34.585H}, celestial mechanics
\citep{2012ascl.soft10014L}, quantum electrodynamics \citep{1966Hearn..9.8..573H,1971ACM..14.8..511H,1979CPC..17.1..207B,1996NuPhS..51..142L,2005urhe.book.....G,2017SpriVer..274.2..704F}
or general relativity \citep{1979GReGr..11.6..411F,1977GReGr...8..987C,1977grg..conf...19W,1981GReGr..13...67K}.
Computer algebra for general relativity (GR) has a long history, beginning
almost as early as computer algebra itself in the 1960s.

The first GR program was GEOM, written by J.G. Fletcher \citep{1967ApJ...148L..91F}.
Its main ability was to compute the Riemann tensor of a given metric.
In 1969, R. A. d'Inverno \citep{1969DInverno..12.2.124} developed
ALAM (for Atlas Lisp Algebraic Manipulator) \citep{1975GReGr...6..567D}
and used it to compute the Riemann and Ricci tensors of the Bondi
metric. According to \citep{2018GourgoulhonE}, Bondi and his collaborators
took 6 months to complete the original calculations, while the ALAM
calculation took 4 minutes and resulted in the discovery of 6 errors
in the original paper. Since then, numerous packages have been developed
and various investigations have been carried out using manipulative
algebraic software (CAS) and in many cases the calculations were of
such length that they would have been prohibitively expensive to complete
without the aid of a computer \citep{2018LRR....21....6M}. Its main
advantage is the ability to handle a large number of algebraic calculations
and this particularity has allowed advances in fields of theoretical
physics such as GR or High Energy Physics (HEP).

Many of these problems have used some free, open source, general purpose
software with an emphasis on tensor calculus for GR, such as Java-based
REDBERRY \citep{2013arXiv1302.1219B,2015JPhCS.608a2060P} , SAGE \citep{2018SIAM..10.125Z}
and SAGEMANIFOLDS written in Python, CADABRA\citep{200CPhCo..176.8..550P,2010CPhCo.181..489B}
developed in C++ and Python. Others, applied specifically for HEP,
have been designed based on special algorithms \citep{1980CoPhC..20...85G,1980SvPhU..23...59G,2017SpriVer..274.2..704F}
and implemented in programs like SCHOONSHIP \citep{1979CoPhC..18....1S,1991ARXIV..6228.v2.10V},
designed by M. Veltman, ASHMEDAI \citep{1967JCPhy1..454L,1974PhRvD...9..421L,1976PACMS..S76..359L}
by M. Levine, REDUCE \citep{1966Hearn..9.8..573H,1971ACM..14.8..511H,1979CPC..17.1..207B,2005urhe.book.....G}
by A. Hearn, MACSYMA (to become MAXIMA in 1998) \citep{1979MITLCS....671L,1985JSC..1..100P,2012JSC..47..130M}
by J. Moses developed at MIT, or more recent FORM \citep{2015NPPP..261...45S,2017arXiv170706453R,2017rcaq.confE..22V}
by J. Vermaseren. Proprietary software such as MATHEMATICA \citep{1980PhDT........69W,1985PhRvL..54..735W,2019APS..MARF22001W,2020arXiv200408210W}by
S. Wolfram or MAPLE \citep{1983SpringerP..83..101C,1997CoPhC..105.2..233P,2000physics..10053V,2001gr.qc.....3023K,2002gr.qc.....9096V,2003JSC..35.3..305S,2004cs........9006V,2006WileyVCH..122622W,2014arXiv1411.7969S}
from B. Char developed at the University of Waterloo. The CAS programs
mentioned above constitute only a very small part of the available
applications, special purpose and general systems that can be consulted
in repositories and lists, maintained and frequently updated \citealp{2021SWMATH,2021SIGSAM,2021Maxima,2021RISC,2021FOLDOC,2021CAN}.
Of all the CAS applications mentioned above, we are going to refer
in this work to \textbf{MAPLE}.

Maple is a general purpose CAS, initially developed at the University
of Waterloo as a result of discussions on the state of symbolic computing
in the 1980s. At the time, large systems such as ALTRAN \citep{1966ACM.1.501B},
CAMAL \citep{1979GReGr..11.6..411F,1977GReGr...8..987C,1977grg..conf...19W},
REDUCE \citep{19733CACMTP..HAV1} and MACSYMA \citep{1979MITLCS....671L},
based on the computer technology of the 1960s, they decided to design
a new system from scratch, taking advantage of the advances in software
engineering available and the lessons of experience. Its basic design
features (for example, \citealp{2013ACMCCA.46.3.164M} elementary
data structures, input/output, number arithmetic, and elementary simplification)
are encoded in a low-level language for efficiency. An important property
is that most of the algebraic facilities of the system are implemented
using the high-level user language. The basic system, or kernel, is
compact and efficient enough to be practical to use for a shared environment
or on personal computers with very little main memory. Library functions
are loaded into the system as needed, adding features such as polynomial
factorization, equation solving, indefinite integration, and matrix
manipulation to the system. The modularity of this design allows users
to demand available computing resources in proportion to their actual
use. It has specialized libraries for elementary and special mathematical
functions, it offers support for symbolic and numerical computation
with exact results, it can handle a wide set of equation systems,
including Diophantine equations, ordinary differential equations (ODE),
partial differential equations (PDE), Differential Algebraic Equations
(DAE), Delay Differential Algebraic Equations (DDE) and recurrence
relations. Initially the kernel of the system was written in macros
that could be translated by a locally developed macroprocessor (called
Margay) into versions of the kernel in the C programming language
for various operating systems; currently only C is used. The GUI was
first released with version 5 (Maple V) and continued to be numbered
until Maple 18 and then changed to a yearly label. Due to the low
demand for main memory to run the kernel and the modular design for
many of the possible user applications in a high-level language, it
has become one of the main symbolic algebra systems used by many researchers
and engineering corporations. worldwide \citep{1991SprUS.1.274C,1992Spr..1..603G}.
These applications or packages can be programmed by the user and many
can be found on the web \citealp{2021MAPLEPKG} as \emph{GRTensorII}
or its update \emph{GRTensor III} . (in short \emph{GRTensor}).

\emph{GRTensor }is a computer algebra package for performing calculations
in the general area of differential geometry \citealp{2021GRTensor3}.
Its main objective is the computation of tensor components in curved
spacetimes specified in terms of a metric or set of basis vectors.
The library relies on a series of special commands starting with $\mathit{"gr"}$
(for example, $\mathit{grcalc,grdisplay,gralter,grdefine}$, etc.)
to deal with a series of (pre) geometric objects. defined as the metric
tensor, Ricci tensor and scalar, Einstein tensor, Chrisstoffell symbols,
etc. This library of objects can be extended to define new tensors,
or use the Newman-Penrose formalism. Although originally designed
for use in the field of general relativity, GRTensor is useful in
many other fields \citealp{1998Musgrave..1.1..22M} There is a version
for \textbf{MATHEMATICA} called \emph{GRTensorM}. The \emph{GRTensorII}
package was originally developed for \emph{Maple V} and can be run
with versions from \emph{Maple V Release 3} to \emph{Maple 13}. The\emph{
GRTensor III} version runs as of \emph{Maple 15}. All documentation
and software are distributed free of charge to help both for research
and teaching.

The final objective of this article is to present the calculations,
especially in Sec. \ref{sec:formbondi} that have been carried out
with an emphasis on the methods, packages and techniques of computational
algebra that we use in a spreadsheet developed for the \emph{GRTensorIII}
package running on the \emph{Maple 17} platform. Initially, it was
written in \emph{GRTensorII} running on top of \emph{Maple 13} and
when upgrading \emph{GRTensor} to version III, it became necessary
to upgrade the \emph{Maple} version as well. The spreadsheet follows
the algorithm described in \citealp{2018PPhy..14.1..46M,2022InJPh..96..317M}
starting from a spherically symmetric perfect fluid distribution of
density $\hat{\rho}$, radial pressure $\hat{P}$, tangential pressure
$\hat{P}_{t}$, a flux of unpolarized radiation moving in the radial
direction with density $\hat{\varepsilon}$ and $3\hat{\mu}$ the
isotropic radiation of the energy density, in a local Minkowskian
system. By performing the corresponding coordinate transformations,
to a radiative (Bondi) coordinate system, we can construct an Impulse
Energy Tensor that must satisfy both the Einstein field equations
$\left(G_{ab}=\kappa T_{ab}\right)$ as the conservation equations
$\left(T_{\:;\,b}^{ab}=0\right)$. When calculating these equations,
it is necessary to simplify them and for this, some \citealp{1998Pertutils},
procedures and functions have been used, as in \citealp{1998gr.qc....10056D}.
In Sec. \ref{sec:Results} some comments on the calculations made
are presented and in the last section the conclusions and the possibility
of extending this procedure to other astrophysical scenarios are detailed.

\section{\label{sec:formbondi}Algorithm structure}

As indicated in the previous section, the created spreadsheet basically
follows the procedure used in \citep{1980PhRvD..22.2305H} and for
the charged case in \citep{1988CaJPh..66..981M} with the modifications
made in \citealp{2018PPhy..14.1..46M,2022InJPh..96..317M}. In the
original sheet, the commands are grouped by sections and each of the
sections has a name that suggests the calculation to be carried out.
In this build for the article, commands (\textcolor{red}{in red})
and output (\textcolor{blue}{in blue}) from the Maple interpreter
will be reproduced. All the commands to perform the calculations of
the sheet are present in this work, however, not all the outputs (\textcolor{blue}{in
blue}) were placed and the most extensive ones were edited.

\subsection{Login }

To start the Maple session we loaded the \emph{GRTensorIII} libraries,
the expression simplification routines were taken from Davies \citep{1998gr.qc....10056D}
and the directory from where the metrics will be loaded. We start
with a reset statement to ensure that you have a new session of the
sheet, 
\begin{lyxcode}
\textcolor{red}{>restart;}

\textcolor{red}{>grtw();}
\end{lyxcode}
\textcolor{blue}{
\begin{equation}
\begin{array}[b]{c}
"\\
\mathit{GRTensorIII\:v2.2\:Oct\,1,\:2018}"\\
"\mathit{Copyright\:2018,\:Peter\:Musgrave,\:Denis\:Pollney,\:Kayll\:Lake}"\\
\mathit{"Latest\:version\:is\:at\:http://github.com/grtensor/grtensor"}\\
\mathit{"For\:help\:?grtensor"}\\
\mathit{"Support/contact\:grtensor3@gmail.com"}\\
\mathit{\boldsymbol{module()\:...end\:module()}}
\end{array}\label{eq:grtw}
\end{equation}
}
\begin{lyxcode}
\textcolor{red}{>with(grtensor):~groptions():~}
\end{lyxcode}
The expression simplification routines were taken from Davies \citep{1998gr.qc....10056D}
\begin{lyxcode}
\textbf{\textcolor{red}{>read~`/home/victor/relatividad/maple/loadmpl/pertutils.mpl`;}}
\end{lyxcode}

\subsection{Minkowskian local flat space}

It is assumed that the region of the space to be considered is composed
of energy density anisotropic material $\hat{\rho}$, radial pressure$\hat{P}$,
tangential pressure $\hat{P}_{t}$, radiation isotropic energy density
$3\hat{\mu}$ and non-polarized radiation of energy density $\hat{\xi}$
propagating in a radial direction. We are going to establish the metric
of this region of space. 
\begin{lyxcode}
\textbf{\textcolor{red}{>qload(minkowski);}}

\textcolor{blue}{Calculated~ds~for~minkowski~(0.001000~sec.)}
\end{lyxcode}
\textcolor{blue}{
\begin{equation}
\begin{array}[b]{c}
\mathit{Default\,spacetime=minkowski}\\
\mathit{For\,the\,minkowski\,spacetime:}\\
\mathit{Coordinates}\\
\mathit{x(up)}\\
\mathit{x^{\alpha}=\left[t\;x\;y\;z\right]}\\
\mathit{Line\ element}\\
\mathit{ds^{2}=dt^{2}-dx^{2}-dy^{2}-dz^{2}}\\
\mathit{The\ Minkowski\:metric-plana-}
\end{array}
\end{equation}
}The covariant components of the metric tensor of the local Minkowski
system are calculated and displayed. You can use two \textbf{\emph{GRTensorIII}}
instructions: $\mathit{grcalc()}$ to calculate and $\mathit{grdisplay()}$
to show the result. However, $\mathit{grcalcd()}$, allows you to
do the calculation and shows the result, combining the two previous
instructions. 
\begin{lyxcode}
\textcolor{red}{>grcalcd(g(dn,dn));}
\end{lyxcode}
Let's set the values for the contravariant unit vectors in the Minkowski
reference system. We will use them to calculate the components of
the  tensor in the local system 
\begin{lyxcode}
\textbf{\textcolor{red}{>}}\textcolor{red}{grdef(`u0\{\textasciicircum a~\}:=kdelta\{\$t~\textasciicircum a\}~`):~grdef(`u1\{\textasciicircum a~\}:=kdelta\{\$x~\textasciicircum a\}~`):}

\textcolor{red}{{}~grdef(`u2\{\textasciicircum a~\}:=kdelta\{\$y~\textasciicircum a\}~`):~grdef(`u3\{\textasciicircum a~\}:=kdelta\{\$z~\textasciicircum a\}~`):}
\end{lyxcode}
And to establish the Stress-energy tensor of this region, let's start
with: Energy density anisotropic material $\rho$, Radial pressure
$P$ and Tangential pressure $P_{t}$
\begin{lyxcode}
\textcolor{red}{>grdef(`TM\{\textasciicircum a~b\}:=~(rho\_M~+~P\_t){*}u0\{\textasciicircum a~\}{*}u0\{b~\}-~P\_t~{*}g\{\textasciicircum a~b\}~+~(P\_M~-~P\_t){*}u1\{\textasciicircum a\}{*}u1\{b\}~`):}

\textbf{\textcolor{red}{>grcalcd(TM(dn,~dn));}}
\end{lyxcode}
Isotropic radiation of energy density $3\mu$
\begin{lyxcode}
\textcolor{red}{>grdef(`Tp\{\textasciicircum a~b\}:=~3{*}u0\{\textasciicircum a~\}{*}u0\{~b\}~+u1\{\textasciicircum a~\}{*}u1\{~b\}+u2\{\textasciicircum a~\}{*}u2\{~b\}+u3\{\textasciicircum a~\}{*}u3\{~b\}`);}
\end{lyxcode}
The component corresponding to Polarized radiation
\begin{lyxcode}
\textcolor{red}{>grcalcd(Tp(dn,~dn));}
\end{lyxcode}
Unpolarized radiation of energy density $\xi$ propagating in the
radial direction
\begin{lyxcode}
\textcolor{red}{>~grdef(`~v\{a~\}:={[}1,~-1,~0,~0~{]}`);~}

\textcolor{red}{>~grdef(`Tnp\{a~b\}:=~xi~{*}v\{a~\}{*}v\{b~\}~`);~}

\textcolor{red}{>~grcalcd(Tnp(dn,~dn));~}
\end{lyxcode}
Bringing together the three parts of the Stress-Energy tensor, we
get the stress-energy tensor Matter + radiation:
\begin{lyxcode}
\textcolor{red}{>~grdef(`T0\{a~b\}:=TM\{a~b\}+mu{*}Tp\{a~b\}+Tnp\{a~b\}`);~}

\textcolor{red}{>~grcalcd(T0(dn,~dn));~}

\textcolor{blue}{Calculated~T0(dn,dn)~for~minkowski~(0.003000~sec.)~}
\end{lyxcode}
\textcolor{blue}{
\[
\begin{array}[t]{c}
\mathit{\mathit{\mathit{`CPUTime`=0.005}}}\\
\mathit{For\,the\,minkowski\,spacetime:}\\
\mathit{T0(dn,dn)}\\
\mathit{T0(dn,dn)}
\end{array}
\]
\begin{equation}
T0_{ab}=\left[\begin{array}{cccc}
3\mu+\mathit{rho_{M}}+\xi & -\xi & 0 & 0\\
-\xi & P_{M}+\mu+\xi & 0 & 0\\
0 & 0 & P_{t}+\mu & 0\\
0 & 0 & 0 & P_{t}+\mu
\end{array}\right]\label{eq:T0dndn}
\end{equation}
}Redefining the radiation and matter variables of the stress-energy
tensor:
\begin{lyxcode}
\textcolor{red}{>~grmap(T0(dn,~dn),~subs,~rho\_M~=~rho{[}0{]}-3{*}mu,~`x`);}

\textcolor{red}{>~grmap(T0(dn,~dn),~subs,~P\_M~=~P{[}0{]}-mu,~`x`);~}

\textcolor{red}{>~grmap(T0(dn,~dn),~subs,~P\_t~=~P{[}t{]}-mu,~`x`);~}

\textcolor{red}{>~grdisplay(T0(dn,dn));}~~~~~
\end{lyxcode}
\textcolor{blue}{
\[
\begin{array}[t]{c}
\mathit{For\,the\,minkowski\,spacetime:}\\
\mathit{T0(dn,dn)}\\
\mathit{T0(dn,dn)}
\end{array}
\]
\begin{equation}
T0_{a\,b}=\left[\begin{array}{cccc}
\rho_{0}+\xi & -\xi & 0 & 0\\
-\xi & P_{0}+\xi & 0 & 0\\
0 & 0 & P_{t} & 0\\
0 & 0 & 0 & P_{t}
\end{array}\right]\label{eq:TOdndn}
\end{equation}
}

\subsection{Lorentz transformation}

Let's suppose that you have an observer moving in relation to the
local system Minkowskian, with a radial velocity $\omega$. The components
of the Stress-Energy tensor in this new system of Lorentz, will be
given by the relationship: 

\[
\bar{T}_{\mu\nu}=\Lambda_{\:\mu}^{\alpha}\Lambda_{\:\nu}^{\beta}\hat{T}_{\alpha\beta}
\]
where $\Lambda_{\:\mu}^{\alpha}$ is the transformation matrix 
\begin{lyxcode}
\textcolor{red}{>~grdef(`umsqrto~:=~1/sqrt(1-omega\textasciicircum 2)~`);}~

\textcolor{red}{grdef(`Lambda\{\textasciicircum a~b\}:=~umsqrto~{*}kdelta\{\textasciicircum a~\$t\}{*}kdelta\{\$t~b\}~}

\textcolor{red}{-omega{*}~umsqrto~{*}kdelta\{\textasciicircum a~\$x\}{*}~kdelta\{\$t~b\}~-omega{*}~umsqrto~{*}kdelta\{\textasciicircum a~\$t\}{*}~kdelta\{\$x~b\}}

\textcolor{red}{+~umsqrto~{*}kdelta\{\textasciicircum a~\$x\}{*}~kdelta\{\$x~b\}+~kdelta\{\textasciicircum a~\$y\}{*}~kdelta\{\$y~b\}+}

\textcolor{red}{kdelta\{\textasciicircum a~\$z\}{*}~kdelta\{\$z~b\}`);}

\textcolor{red}{>~grcalcd(Lambda(up,~dn));}

\end{lyxcode}
\textcolor{blue}{
\[
\begin{array}[t]{c}
\mathit{CPU\:Time=0.011}\\
\mathit{\mathit{For\,the\,minkowski\,spacetime:}}\\
\mathit{Lambda(up,dn)}\\
\mathit{\Lambda(up,dn)}
\end{array}
\]
\begin{equation}
\Lambda_{\:\mu}^{\alpha}={\color{blue}}\left[\begin{array}{cccc}
\frac{1}{\sqrt{1-\omega^{2}}} & -\frac{\omega}{\sqrt{1-\omega^{2}}} & 0 & 0\\
-\frac{\omega}{\sqrt{1-\omega^{2}}} & \frac{1}{\sqrt{1-\omega^{2}}} & 0 & 0\\
0 & 0 & 1 & 0\\
0 & 0 & 0 & 1
\end{array}\right],\label{eq:MLambdaupdn}
\end{equation}
}Let's carry out the transformation operation now. To do it, we will
establish this transformation as:
\begin{lyxcode}
\textcolor{red}{>~grdef(`T1\{a~b\}:=Lambda\{\textasciicircum c~a\}{*}Lambda\{\textasciicircum d~b\}{*}T0\{c~d\}`);}

\textcolor{red}{>~grcalcd(T1(dn,~dn));~}\textcolor{blue}{
\[
\begin{array}[t]{c}
\mathit{CPU\:Time=0.012}\\
\mathit{For\,the\,minkowski\,spacetime:}\\
\mathit{T1\left(dn,dn\right)}\\
\mathit{\Lambda(up,dn)}
\end{array}
\]
\begin{equation}
T1_{a\,b}=\left[\begin{array}{cccc}
-\frac{\rho_{0}+2\omega\xi+\omega^{2}P_{0}+\omega^{2}\xi}{-1+\omega^{2}} & \:\frac{\omega_{0}\rho_{0}+2\omega\xi+\xi+\omega P_{0}}{-1+\omega^{2}} & 0 & 0\\
\frac{\omega_{0}\rho_{0}+2\omega\xi+\xi+\omega P_{0}}{-1+\omega^{2}} & -\frac{\omega^{2}\rho_{0}+\omega^{2}\xi+2\omega\xi+P_{0}+\xi}{-1+\omega^{2}} & 0 & 0\\
0 & 0 & P_{t} & 0\\
0 & 0 & 0 & P_{t}
\end{array}\right]\label{eq:T1dndn-1}
\end{equation}
}
\end{lyxcode}
We obtain the expression of the Stress-energy tensor in the local
system, with radial velocity $\omega$.

\subsection{\label{subsec:BRiCS}Bondi Radiative Coordinate System }

As the study we are doing, is related to radiation, then it is logical
to assume that we must use a coordinate system according to the theme.
Therefore, we are going to use Bondi's radiative coordinate system
as in \citep{1964RSPSA.281...39B}. 
\begin{lyxcode}
\textcolor{red}{>qload(bondi);~}

\textcolor{blue}{Calculated~ds~for~bondi~(0.001000~sec.)~}
\end{lyxcode}
\textcolor{blue}{
\begin{equation}
\begin{array}[b]{c}
\mathit{Default\,space\,time=bondi}\\
\mathit{For\,the\,bondi\,spacetime:}\\
\mathit{Coordinates}\\
\mathit{x\left(up\right)}\\
\mathit{x^{\alpha}=\bar{\left[u,r,\theta,\phi\right]}}\\
\mathit{Line\:elemen}t\\
\mathit{ds^{2}=\frac{V\left(u,r\right)\cdot e^{2\beta\left(u,r\right)}}{r}du^{2}+2e^{2\beta\left(u,r\right)}dr\cdot du-r^{2}d\theta^{2}-r^{2}\sin^{2}\theta d\phi^{2}}\\
\mathit{The\,Bondi\,metric\,(Proc.\,Roy.\,Soc.\,A\,269\;21)}
\end{array}\label{eq:DefBondiMet}
\end{equation}
}Changing the expression of the Stress-energy tensor of the local
Minkowskiana metric to its structure in the Bondi radiation coordinate
system. 
\begin{lyxcode}
\textcolor{red}{>grdef(`Vsr:=sqrt(V(u,r)/r)`);~}

\textcolor{red}{>grdef(`Mu\{\textasciicircum a~b\}:=~exp(beta(u,r)){*}Vsr{*}kdelta\{\textasciicircum a~\$u\}{*}kdelta\{\$u~b\}+}

\textcolor{red}{exp(beta(u,r))/Vsr{*}kdelta\{\textasciicircum a~\$u\}{*}kdelta\{\$r~b\}+~exp(beta(u,r))/}

\textcolor{red}{Vsr{*}kdelta\{\textasciicircum a~\$r\}{*}kdelta\{\$r~b\}+r{*}kdelta\{\textasciicircum a~\$theta\}{*}kdelta\{\$theta~b\}+}

\textcolor{red}{r{*}sin(theta){*}kdelta\{\textasciicircum a~\$phi\}{*}kdelta\{\$phi~b\}`);~}
\end{lyxcode}
Defining and showing the matrix of the Minkowski local system transformation
to the Lorentz system
\begin{lyxcode}
\textcolor{red}{>grcalcd(Mu(up,~dn));~}

\textcolor{blue}{Calculated~Vsr~for~bondi~(0.002000~sec.)~}

\textcolor{blue}{Calculated~grtensor:-kdelta(dn,dn)~for~bondi~(0.002000~sec.)~}

\textcolor{blue}{Calculated~grtensor:-kdelta(up,dn)~for~bondi~(0.002000~sec.)~}

\textcolor{blue}{Calculated~Mu(up,dn)~for~bondi~(0.011000~sec.)~}

\textcolor{blue}{
\[
\begin{array}[t]{c}
\mathit{CPU\,Time=0.021}\\
\mathit{For\,the\,bondi\,spacetime:}\\
\mathit{Mu\left(up,dn\right)}\\
\mathit{M\left(up,dn\right)}
\end{array}
\]
}
\end{lyxcode}
\textcolor{blue}{
\begin{equation}
M_{\hspace*{1em}b}^{a}=\left[\begin{array}{cccc}
e^{2\beta\left(u,r\right)}\sqrt{\frac{V\left(u,r\right)}{r}} & \frac{e^{2\beta\left(u,r\right)}}{\sqrt{\frac{V\left(u,r\right)}{r}}} & 0 & 0\\
0 & \frac{e^{2\beta\left(u,r\right)}}{\sqrt{\frac{V\left(u,r\right)}{r}}} & 0 & 0\\
0 & 0 & r & 0\\
0 & 0 & 0 & r\sin\theta
\end{array}\right]\label{eq:Mink2Bond_Matrix}
\end{equation}
}This matrix of transformation, allows us to obtain the expression
of the stress-energy tensor of the local system of Minkowski to the
system of radiative coordinates with

\begin{equation}
\bar{T}_{ab}=M_{\hspace*{1em}a}^{\alpha}M_{\hspace*{1em}b}^{\beta}\hat{T}_{\:\alpha\beta}\label{eq:Mink2Bondi}
\end{equation}
where $M_{\ b}^{a}$ is the transformation matrix. As we are performing
an operation between two expressions with different metrics, we must
specify the space - or metric - corresponding to each term of the
multiplication 
\begin{lyxcode}
\textcolor{red}{>~grdef(`TB<2>\{a~b\}:=Mu\{\textasciicircum c~a\}~{*}Mu\{\textasciicircum d~b\}{*}T1<1>\{c~d\}`);}
\end{lyxcode}
by defining what is the scope of the definition of each term
\begin{lyxcode}
\textcolor{red}{>~grcalcd(1~=~minkowski,~2~=~bondi,~TB(dn,~dn));}\textcolor{blue}{
\[
\begin{array}[t]{c}
\mathit{CPU\:Time=0.202}\\
\mathit{For\,the\,bondi\,spacetime:}\\
\mathit{TB\left(dn,dn\right)}\\
\mathit{TB\left(dn,dn\right)}
\end{array}
\]
}
\end{lyxcode}
\textcolor{blue}{
\begin{equation}
TB_{ab}=\left[\begin{array}{cccc}
-\frac{e^{2\beta\left(u,r)\right)}V\left(u,r\right)\left(\rho_{0}+2\omega\xi+\omega^{2}P_{0}+\omega^{2}\xi\right)}{r\left(-1+\omega^{2}\right)} & -\frac{\left(\omega P_{0}-\rho_{0}\right)}{\omega+1}e^{2\beta\left(u,r\right)} & 0 & 0\\
-\frac{\left(\omega P_{0}-\rho_{0}\right)}{\omega+1}e^{2\beta\left(u,r\right)} & -\frac{\left(\omega\rho_{0}+\omega P_{0}-\rho_{0}-P_{0}\right)r}{V\left(u,r\right)\left(\omega+1\right)} & 0 & 0\\
0 & 0 & r^{2}P_{t} & 0\\
0 & 0 & 0 & r^{2}\sin^{2}\theta\,P_{t}
\end{array}\right]\label{eq:TBdndn}
\end{equation}
}

\subsection{Effective Variables}

We are going to define the effective variables in a similar way to
the case of \citep{1980PhRvD..22.2305H,2018PPhy..14.1..46M}. However,
here we will denote it as $\rho{}_{1}$ and $P_{1}$, without a bar
at the top: 
\begin{lyxcode}
\textcolor{red}{>~grmap(TB(dn,~dn),~subs,~rho{[}0{]}~=~(rho{[}1{]}+omega{*}P{[}1{]})/(1-omega),~`x`);~}

\textcolor{red}{>~grmap(TB(dn,~dn),~subs,~P{[}0{]}~=~(P{[}1{]}+omega{*}rho{[}1{]})/(1-omega),~`x`);~}
\end{lyxcode}
Changing the term a bit for radiation
\begin{lyxcode}
\textcolor{red}{>~grmap(TB(dn,~dn),~subs,~xi~=~(1-omega){*}epsilon/(1+omega),~`x`);}

\textcolor{red}{>~gralter(TB(dn,~dn),~expand,~factor);~}

\textcolor{red}{>~grdisplay(TB(dn,~dn));~}
\end{lyxcode}
\textcolor{blue}{
\[
\begin{array}[t]{c}
\mathit{For\,the\,bondi\,spacetime:}\\
\mathit{TB\left(dn,dn\right)}\\
\mathit{TB\left(dn,dn\right)}
\end{array}
\]
}

\textcolor{blue}{
\begin{equation}
TB_{ab}=\left[\begin{array}{cccc}
\frac{e^{2\beta\left(u,r)\right)}V\left(u,r\right)\left(\omega^{2}\varepsilon+\omega^{2}\rho_{1}+\omega P_{1}-2\varepsilon\omega-\omega\rho_{1}+\varepsilon+\rho_{1}\right)}{r\left(-1+\omega\right)^{2}} & \rho_{1}\,e^{2\beta\left(u,r\right)} & 0 & 0\\
\rho_{1}\,e^{2\beta\left(u,r\right)} & \frac{re^{2\beta\left(u,r)\right)}\left(\rho_{1}+P_{1}\right)}{V\left(u,r\right)} & 0 & 0\\
0 & 0 & r^{2}P_{t} & 0\\
0 & 0 & 0 & r^{2}\sin^{2}\theta\,P_{t}
\end{array}\right]\label{eq:TBdndn-1}
\end{equation}
}Using the routines defined in \citep{1998gr.qc....10056D} we can
obtain an alternate expression for $TB_{uu}$:
\begin{lyxcode}
\textcolor{red}{>~TBdndnuu~:=~kfactor(hcollect(grcomponent(TB(dn,~dn),~{[}u,~u{]}),~\{P{[}1{]},~rho{[}1{]},~V(u,~r)\},}

\textcolor{red}{\{r,~omega,~beta(u,~r)\}),~V(u,~r){*}exp(2{*}beta(u,~r))/r);~}
\end{lyxcode}
\textcolor{blue}{
\begin{eqnarray}
\mathit{TBdndnuu} & \mathit{=} & \mathit{\frac{V\left(u,r\right)}{r}e^{2\beta\left(u,r\right)}\left(\frac{\omega P_{1}}{\left(-1+\omega\right)^{2}}+\frac{\left(-\omega+\omega^{2}+1\right)\rho_{1}}{\left(-1+\omega\right)^{2}}+\varepsilon\right)}
\end{eqnarray}
}or equivalently
\[
\mathit{TBdndnuu=\frac{V\left(u,r\right)}{r}e^{2\beta\left(u,r\right)}\left[\frac{\omega\left(\rho_{1}+P_{1}\right)}{\left(1-\omega\right)^{2}}+\rho_{1}+\varepsilon\right]}
\]

\subsection{Electromagnetic Component}

If we are interested in the possibility that the material subject
to the study presents an electric charge, it is necessary to include
it in the tensor and therefore establish the expression for the Faraday
tensor 
\begin{equation}
T_{\alpha\beta}=\frac{1}{4\pi}\left[F_{\alpha\mu}F_{\negmedspace\beta}^{\mu}+\frac{1}{4}g_{\alpha\beta}\left(F_{\mu\nu}F^{\mu\nu}\right)\right]\label{eq:Tem}
\end{equation}
where $F^{ab}$satisfies the equations
\[
\left(\sqrt{g}F^{\alpha\beta}\right)_{,\beta}=4\pi\sqrt{g}\,J^{\alpha}\qquad F_{\left[\alpha\beta;\gamma\right]}=0
\]
Assuming spherical symmetry, the only non-zero component is $F$ and
we have $F^{01}$ 
\begin{eqnarray*}
\left(e^{2\beta}r^{2}F^{01}\right)_{,1} & = & 4\pi e^{2\beta}r^{2}J^{0}\\
\left(e^{2\beta}r^{2}F^{10}\right)_{,0} & = & 4\pi e^{2\beta}r^{2}J^{1}
\end{eqnarray*}
and when integrating the first:

\[
\left(e^{2\beta}r^{2}F^{01}\right)=\int_{0}^{r}\left(4\pi e^{2\beta}r^{2}J^{0}\right)dr\equiv Q\left(u,r\right)
\]
Because of this, we can write
\[
F^{01}=\frac{e^{-2\beta}}{r^{2}}\cdot Q\left(u,r\right)
\]
where $Q\left(u,r\right)$, plays the role of electric charge and
the electromagnetic component of the stress tensor is 
\begin{lyxcode}
\textcolor{red}{>~grdef(`F\{\textasciicircum a~\textasciicircum b\}:=~F01{*}(kdelta\{\$u~\textasciicircum a\}{*}kdelta\{\$r~\textasciicircum b\}~-kdelta\{\$r~\textasciicircum a\}{*}kdelta\{\$u~\textasciicircum b\})`);~}

\textcolor{red}{>~grdef(`Faraday2:=F\{\textasciicircum a~\textasciicircum b\}{*}F\{a~b\}~`);~}

\textcolor{red}{>~grdef(`Tem\{a~b\}:=~(F\{a~c\}{*}~F\{\textasciicircum c~b\}+(1/4){*}g\{a~b\}{*}F\{c~d\}{*}F\{\textasciicircum c~\textasciicircum d~\})/(4{*}Pi)~`);}

\textcolor{red}{>~grcalcd(Faraday2);}

\textcolor{red}{>~grcalcd(F(up,~up));}
\end{lyxcode}
\textcolor{blue}{
\[
\begin{array}[t]{c}
\mathit{CPU\,Time=0}.\\
\mathit{For\,the\,bondi\,spacetime:}\\
\mathit{F(up,up)}\\
\mathit{F(up,up)}
\end{array}
\]
\begin{equation}
F^{ab}=\left[\begin{array}{cccc}
0 & F01 & 0 & 0\\
-F01 & 0 & 0 & 0\\
0 & 0 & 0 & 0\\
0 & 0 & 0 & 0
\end{array}\right],\label{eq:Fupup}
\end{equation}
}In terms of $F^{01}$, the electromagnetic component of the stress
tensor is
\begin{lyxcode}
\textcolor{red}{>~grcalcd(Tem(dn,~dn));~}

\textcolor{blue}{Calculated~g(up,up)~for~bondi~(0.004000~sec.)~}

\textcolor{blue}{Calculated~F(up,dn)~for~bondi~(0.004000~sec.)~}

\textcolor{blue}{Calculated~Tem(dn,dn)~for~bondi~(0.000000~sec.)~}

\end{lyxcode}
\textcolor{blue}{
\[
\begin{array}[t]{c}
\mathit{CPU\,Time=0.011}\\
\mathit{For\,the\,bondi\,spacetime:}\\
\mathit{Tem(dn,dn)}\\
\mathit{Tem(dn,dn)}
\end{array}
\]
\begin{equation}
Tem_{ab}=\left[\begin{array}{cccc}
\frac{1}{8}\frac{V\left(u,r\right)\left(e^{2\beta\left(u,r\right)}\right)^{3}F01^{2}}{\pi r} & \frac{1}{8}\frac{\left(e^{2\beta\left(u,r\right)}\right)^{3}F01^{2}}{\pi} & 0 & 0\\
\frac{1}{8}\frac{\left(e^{2\beta\left(u,r\right)}\right)^{3}F01^{2}}{\pi} & 0 & 0 & 0\\
0{\color{blue}} & 0 & \frac{1}{8}\frac{r^{2}\left(e^{2\beta\left(u,r\right)}\right)^{3}F01^{2}}{\pi} & 0\\
0 & 0 & 0 & \frac{1}{8}\frac{r^{2}\sin\left(\theta\right)^{2}\left(e^{2\beta\left(u,r\right)}\right)^{3}F01^{2}}{\pi}
\end{array}\right],
\end{equation}
}The electromagnetic component of the stress tensor is expressed in
terms of the electric charge as:
\begin{lyxcode}
\textcolor{red}{>~grmap(Tem(dn,~dn),~subs,~F01~=~exp(-2{*}beta(u,~r)){*}Q(u,~r)/r\textasciicircum 2,~`x`);~}

\textcolor{red}{>~grmap(Faraday2,~subs,~F01~=~exp(-2{*}beta(u,~r)){*}Q(u,~r)/r\textasciicircum 2,~`x`);}

\textcolor{red}{>~gralter(Tem(dn,~dn),~expand,~factor);~}

\textcolor{blue}{Component~simplification~of~a~GRTensorIII~object:}

\textcolor{blue}{Applying~routine~expand~to~object~Tem(dn,dn)~}

\textcolor{blue}{Applying~routine~factor~to~object~Tem(dn,dn)~}
\end{lyxcode}
\textcolor{blue}{
\[
\mathit{CPUTime=0.007}
\]
}
\begin{lyxcode}

\textcolor{red}{>~grdisplay(Tem(dn,~dn));~}
\end{lyxcode}
\textcolor{blue}{
\[
\begin{array}[t]{c}
\mathit{For\,the\,bondi\,spacetime:}\\
\mathit{Tem(dn,dn)}\\
\mathit{Tem(dn,dn)}
\end{array}
\]
\begin{equation}
\mathit{Tem_{ab}}=\left[\begin{array}{cccc}
\frac{1}{8}\frac{V\left(u,r\right)\left(e^{2\beta\left(u,r\right)}\right)^{2}Q^{2}\left(u,r\right)}{\pi r^{5}} & \frac{1}{8}\frac{\left(e^{2\beta\left(u,r\right)}\right)^{2}Q^{2}\left(u,r\right)}{\pi r^{4}} & 0 & 0\\
\frac{1}{8}\frac{\left(e^{2\beta\left(u,r\right)}\right)^{2}Q^{2}\left(u,r\right)}{\pi r^{4}} & 0 & 0 & 0\\
0 & 0 & \frac{1}{8}\frac{Q^{2}\left(u,r\right)}{\pi r^{2}} & 0\\
0 & 0 & 0 & \frac{1}{8}\frac{\sin\left(\theta\right)^{2}Q^{2}\left(u,r\right)}{\pi r^{2}}
\end{array}\right].
\end{equation}
}

\subsection{Stress tensor }

The stress tensor total that includes matter + radiation + electric
charge is:
\begin{lyxcode}
\textcolor{red}{>~grdef(`T\{a~b\}~:=~TB\{a~b\}+Tem\{a~b\}`);}
\end{lyxcode}
If we want to study neutral or uncharged cases, we can turn OFF the
contribution of the electric charge 
\begin{lyxcode}
\textcolor{red}{>~\#grmap(T(dn,~dn),~subs,~Q(u,~r)~=~0,~`x`)}
\end{lyxcode}
Calculating the covariant components
\begin{lyxcode}
\textcolor{red}{>~grcalc(T(dn,~dn));~}

\textcolor{red}{>~gralter(T(dn,~dn),~expand,~factor);~}

\textcolor{red}{>~grdisplay(T(dn,~dn));~}
\end{lyxcode}
\textcolor{blue}{
\[
\begin{array}[t]{c}
\mathit{For\,the\,bondi\,spacetime:}\\
\mathit{T(dn,dn)}\\
\mathit{T(dn,dn)}
\end{array}
\]
\begin{multline}
\begin{array}[b]{c}
T_{uu}=\frac{1}{8}\frac{1}{(-1+\omega)^{2}r^{5}\pi}\left(V(u,r)\cdot\left(e^{\beta\left(u,r\right)}\right)^{2}\left(8r^{4}\pi\varepsilon\omega^{2}+8r^{4}\pi\omega^{2}\rho_{1}+8r^{4}\pi\omega P_{1}-16r^{4}\pi\varepsilon\omega\right.\right.\\
\left.\left.-8r^{4}\pi\omega\rho_{1}+8r^{4}\pi\varepsilon+8r^{4}\pi\rho_{1}+\left(Q\left(u,r\right)\right)^{2}-2\left(Q\left(u,r\right)\right)^{2}\omega+\omega^{2}\left(Q\left(u,r\right)\right)^{2}\right)\right)\\
T_{ru}=\frac{1}{8}\frac{\left(e^{\beta\left(u,r\right)}\right)^{2}\left(8r^{4}\pi\rho_{1}+Q^{2}\left(u,r\right)\right)}{r^{4}\pi}\\
T_{ur}=\frac{1}{8}\frac{\left(e^{\beta\left(u,r\right)}\right)^{2}\left(8r^{4}\pi\rho_{1}+Q^{2}\left(u,r\right)\right)}{r^{4}\pi}\\
T_{rr}=\frac{r\left(e^{\beta\left(u,r\right)}\right)^{2}(P_{1}+\rho_{1})}{V(u,r)}\\
T_{\theta\theta}=\frac{1}{8}\frac{8r^{4}P_{t}\pi+Q\left(u,r\right)^{2}}{\pi r^{2}}\\
T_{\phi\phi}=\frac{1}{8}\frac{\sin\theta^{2}\left(8r^{4}P_{t}\pi+Q\left(u,r\right)^{2}\right)}{\pi r^{2}}
\end{array}
\end{multline}
}As an example of the use of simplification routines, we can factor
the time covariant component
\begin{lyxcode}
\textcolor{red}{>tempTdndnuu~:=~kfactor(hcollect(grcomponent(T(dn,~dn),~{[}u,~u{]}),~\{P{[}1{]},~rho{[}1{]},~V(u,~r)\},~}

\textcolor{red}{\{r,~omega,~beta(u,~r)\}),~V(u,~r){*}exp(2{*}beta(u,~r))/r);}
\end{lyxcode}
\textcolor{blue}{
\begin{equation}
\mathit{tempTdndnuu:=\frac{V\left(u,r\right)e^{2\beta\left(u,r\right)}\left(\frac{\omega P_{1}}{\left(-1+\omega\right)^{2}}+\frac{\left(-\omega+\omega^{2}+1\right)\rho_{1}}{\left(-1+\omega\right)^{2}}+\frac{1}{8}\left(\frac{8r^{4}\pi\varepsilon+Q\left(u,r\right)^{2}}{r^{4}\pi}\right)\right)}{r}}
\end{equation}
}The contravariant radial component can be simplified with the help
of the previously loaded factoring functions
\begin{lyxcode}
\textcolor{red}{>~tempTupuprr~:=~kfactor(hcollect(hcollect(grcomponent(T(up,~up),~{[}r,~r{]}),~\{epsilon,~}

\textcolor{red}{Q(u,~r),~V(u,~r),~exp(2{*}beta(u,~r))\},~\{epsilon,~P{[}1{]},~rho{[}1{]},~Q(u,~r)\}),~\{epsilon,~}

\textcolor{red}{P{[}1{]},~rho{[}1{]},~Q(u,~r)\},~\{V(u,~r),~beta(u,~r)\}),~exp(-2{*}beta(u,~r)){*}V(u,~r/}r)
\end{lyxcode}
\textcolor{blue}{
\begin{equation}
\mathit{tempTupuprr=\frac{V\left(u,r\right)e^{-2\beta\left(u,r\right)}\left(\varepsilon+\frac{\left(-\omega+\omega^{2}+1\right)P_{1}}{\left(-1+\omega\right)^{2}}-\frac{1}{8}\frac{Q\left(u,r\right)^{2}}{r^{4}\pi}+\frac{\omega\rho_{1}}{\left(-1+\omega\right)^{2}}\right)}{r}}
\end{equation}
}that we can simplify as
\[
\mathit{tempTupuprr=\frac{V\left(u,r\right)e^{-2\beta\left(u,r\right)}}{r}\left[\frac{\omega\left(\rho_{1}+P_{1}\right)}{\left(1-\omega\right)^{2}}+\varepsilon+\left(P_{1}-\frac{Q\left(u,r\right)^{2}}{8\pi r^{4}}\right)\right]}
\]

\subsection{Temporary and radial dependency of the Stress Tensor}

To calculate the conservation equations, it is necessary to establish
that the density $\rho=\rho\left(u,r\right)$, pressure $P=P\left(u,r\right)$,
radiation $\varepsilon=\varepsilon\left(u,r\right)$ and tangential
pressure $P_{t}=P_{t}\left(u,r\right)$ depend on that of the temporal
and radial coordinates.
\begin{lyxcode}
\textcolor{red}{>~grmap(T(up,~dn),~subs,~rho{[}1{]}~=~rho(u,~r),`x`);}

\textcolor{red}{>~grmap(T(up,~dn),~subs,~P{[}1{]}~=~P(u,~r),~`x`);~}

\textcolor{red}{>~grmap(T(up,~dn),~subs,~P{[}t{]}~=~P{[}t{]}(u,~r),~`x`);}

\textcolor{red}{>~grmap(T(up,~dn),~subs,~epsilon~=~epsilon(u,~r),~`x`);~}

\textcolor{red}{>~grmap(T(up,~dn),~subs,~omega~=~omega(u,~r),~`x`);~}

\textcolor{red}{>~gralter(T(up,~dn),~simplify);~}

\textcolor{red}{>~grdisplay(T(up,~dn));~}
\end{lyxcode}
\textcolor{blue}{
\[
\begin{array}[t]{c}
\mathit{For\,the\,bondi\,spacetime:}\\
\mathit{T(up,dn)}\\
\mathit{T(up,dn)}
\end{array}
\]
\begin{align}
T_{\:b}^{a}= & \left[\:\begin{bmatrix}\frac{1}{8}\left(\frac{8r^{4}\pi\rho(u,r)+Q(u,r)^{2}}{r^{4}\pi}\right), & r\left(\frac{P(u,r)+\rho(u,r)}{V\left(u,r\right)}\right), & 0, & 0\end{bmatrix},\right.\nonumber \\
 & \begin{bmatrix}\frac{V\left(u,r\right)}{r}\frac{\left(\varepsilon\left(u,r\right)\omega\left(u,r\right)^{2}+\omega\left(u,r\right)P\left(u,r\right)-2\varepsilon\left(u,r\right)\omega\left(u,r\right)+\omega\left(u,r\right)\rho\left(u,r\right)+\varepsilon\left(u,r\right)\right)}{\left(-1+\omega\left(u,r\right)\right)^{2}}, & \frac{-Q(u,r)^{2}+8r^{4}\pi P(u,r)}{r^{4}\pi} & 0, & 0\end{bmatrix},\nonumber \\
 & \begin{bmatrix}0, & 0, & -\frac{1}{8}\left(\frac{8r^{4}\pi P_{t}(u,r)+Q(u,r)^{2}}{r^{4}\pi}\right), & 0\end{bmatrix},\nonumber \\
 & \left.\begin{bmatrix}0, & 0, & 0, & -\frac{1}{8}\left(\frac{8r^{4}\pi P_{t}(u,r)+Q(u,r)^{2}}{r^{4}\pi}\right)\end{bmatrix}\,\right]
\end{align}
}

\subsection{\label{subsec:(TOV)}Conservation equations (TOV)}

These equations are important since the Einstein field equations are
composed of two structures: One is the geometry of the $G_{ab}$ system
and the other is basically the energy $T_{ab}$ . From the expressions
that we have taken the geometry is linear in the second derivative
and non-linear in the first. However, the impulse energy tensor that
we have used is linear. Presumably, this non-linearity, of the geometrical
part must have a non-linear version in some expression of the impulse
energy tensor. It is therefore necessary to calculate the components
of the conservation equation since it is possible that they contain
the \emph{ non-linear} expression necessary
to correct our choice of the structure of the Stress Tensor. 
\begin{lyxcode}
\textcolor{red}{>~grdef(`cero\{a~\}:={[}0,~0,~0,~0~{]}`);~}

\textcolor{red}{>~grdef(`TOV\{b\}~:=~T\{\textasciicircum a~b;a\}=~cero\{b\}`);~}
\end{lyxcode}
Calculating the covariant components of the conservation equation:
\begin{lyxcode}
\textcolor{red}{>~grcalc(TOV(dn));~}

\textcolor{red}{>~gralter(TOV(dn),~simplify);~}

\textcolor{red}{>~grdisplay(TOV(dn));~}
\end{lyxcode}
\textcolor{blue}{
\[
\begin{array}[t]{c}
\mathit{For\,the\,bondi\,spacetime:}\\
\mathit{TOV(dn)}\\
\mathit{TOV(dn)}
\end{array}
\]
\begin{flalign}
TOV_{u} & =\left(\frac{1}{4}\frac{1}{r^{4}V\left(u,r\right)\left(-1+\omega\left(u,r\right)\right)^{3}}\left(8\left(\omega\left(u,r\right)^{2}\varepsilon\left(u,r\right)+\left(P\left(u,r\right)+\rho\left(u,r\right)-2\cdot\varepsilon\left(u,r\right)\right)\omega\left(u,r\right)+\varepsilon\left(u,r\right)\right)\cdot\right.\right.\nonumber \\
 & V\left(u,r\right)^{2}r^{3}\pi\left(-1+\omega\left(u,r\right)\right)+\cdots=0\\
TOV_{r} & =\left(\frac{1}{4}\frac{1}{r^{4}V\left(u,r\right)^{2}\pi}\left(4r^{5}\pi V\left(u,r\right)\left(\frac{\partial}{\partial u}\rho\left(u,r\right)\right)+4r^{5}\pi V\left(u,r\right)\left(\frac{\partial}{\partial u}\rho\left(u,r\right)\right)-\cdots\right.\cdots\right.\nonumber \\
 & -4r^{5}\pi P\left(u,r\right)r^{4}\pi\left(\frac{\partial}{\partial u}V\left(u,r\right)\right)-4r^{5}\pi\rho\left(u,r\right)+\cdots=0
\end{flalign}
}We can use the simplification routines loaded at the beginning of
the sheet to compare the terms observed in the conservation equations.
The following simplifications allow us to establish that some of the
identities are not really independent: 
\begin{lyxcode}
\textcolor{red}{>TOVn{[}0{]}~:=~hcollect(simplify(exp(2{*}beta(u,~r)){*}lhs(grcomponent(TOV(up),~{[}u{]}))),~}

\textcolor{red}{\{P(u,~r),~rho(u,~r),~diff(P(u,~r),~r),~diff(Q(u,~r),~r),~P{[}t{]}(u,~r)\},~}

\textcolor{red}{\{diff(V(u,~r),~r),~diff(V(u,~r),~u),~diff(beta(u,~r),~r),~diff(beta(u,~r),~u)\});}
\end{lyxcode}
\textcolor{blue}{
\begin{eqnarray}
TOVn_{0} & = & \left(-\frac{1}{2}\frac{\frac{\partial}{\partial r}V\left(u,r\right)}{V\left(u,r\right)}-\frac{r\left(\frac{\partial}{\partial u}V\left(u,r\right)\right)}{V\left(u,r\right)^{2}}-\left(\frac{\partial}{\partial r}\beta\left(u,r\right)\right)+\frac{2r\frac{\partial}{\partial u}\beta\left(u,r\right)}{V\left(u,r\right)}-\frac{3}{2r}\right)P\left(u,r\right)+\nonumber \\
 &  & -\frac{\partial}{\partial r}P\left(u,r\right)+\frac{1}{4}\frac{Q\left(u,r\right)\frac{\partial}{\partial r}Q\left(u,r\right)}{\pi r^{4}}+\left(-\frac{1}{2}\frac{\frac{\partial}{\partial r}V\left(u,r\right)}{V\left(u,r\right)}-\frac{r\left(\frac{\partial}{\partial u}V\left(u,r\right)\right)}{V\left(u,r\right)^{2}}+\right.\nonumber \\
 &  & \left.-\left(\frac{\partial}{\partial r}\beta\left(u,r\right)\right)+\frac{2r\frac{\partial}{\partial u}\beta\left(u,r\right)}{V\left(u,r\right)}+\frac{1}{2r}\right)\rho\left(u,r\right)+\frac{2P_{t}\left(u,r\right)}{r}+\frac{r\left(\frac{\partial}{\partial u}P\left(u,r\right)+\frac{\partial}{\partial u}\rho\left(u,r\right)\right)}{V\left(u,r\right)}
\end{eqnarray}
}
\begin{lyxcode}
\textcolor{red}{>TOVn{[}1{]}~:=~hcollect(hcollect(simplify(r{*}exp(2{*}beta(u,~r)){*}lhs(grcomponent(TOV(up),~}

\textcolor{red}{{[}r{]}))/V(u,~r)),~\{V(u,~r),~beta(u,~r),~diff(V(u,~r),~r),~diff(V(u,~r),~u),~diff(}

\textcolor{red}{beta(u,~r),~r),~diff(beta(u,~r),~u)\},\{diff(epsilon(u,~r),~r),~diff(epsilon(u,~r),~}

\textcolor{red}{u),~diff(omega(u,~r),~r),~diff(omega(u,~r),~u),~epsilon(u,~r),~omega(u,~r)\}),~}

\textcolor{red}{\{P(u,~r),~Q(u,~r),~rho(u,~r),~diff(P(u,~r),~r),~diff(Q(u,~r),~r),~}

\textcolor{red}{diff(Q(u,~r),~u),~diff(epsilon(u,~r),~r),~epsilon(u,~r),~P{[}t{]}(u,~r)\},~\{V(u,~r),~}

\textcolor{red}{beta(u,~r),~diff(V(u,~r),~r),~diff(V(u,~r),~u),~diff(beta(u,~r),~r),~diff(beta}

\textcolor{red}{(u,~r),~u)\});}
\end{lyxcode}
\textcolor{blue}{
\begin{eqnarray}
TOVn_{1} & := & \left(\frac{\left(\left(\omega\left(u,r\right)\right)^{2}+1\right)\frac{\partial}{\partial r}\beta\left(u,r\right)}{\left(-1+\omega\left(u,r\right)\right)^{2}}+\right.\nonumber \\
 &  & +\frac{1}{2}\frac{-2\,r\frac{\partial}{\partial r}\omega\left(u,r\right)+3\,\left(\omega\left(u,r\right)\right)^{3}-3+7\,\omega\left(u,r\right)-7\,\left(\omega\left(u,r\right)\right)^{2}-\cdots}{r\left(-1+\omega\left(u,r\right)\right)^{3}}\nonumber \\
 &  & -\frac{1}{2}\frac{-\left(\frac{\partial}{\partial r}V\left(u,r\right)\right)\left(\omega\left(u,r\right)\right)^{2}+2\,r\left(\frac{\partial}{\partial u}\beta\left(u,r\right)\right)\left(\omega\left(u,r\right)\right)^{2}-\cdots}{V\left(u,r\right)\left(-1+\omega\left(u,r\right)\right)^{2}}+\nonumber \\
 &  & \left.\frac{-\frac{\partial}{\partial r}V\left(u,r\right)+2\,r\frac{\partial}{\partial u}\beta\left(u,r\right)}{V\left(u,r\right)\left(-1+\omega\left(u,r\right)\right)^{2}}+\frac{1}{2}\,\frac{r\frac{\partial}{\partial u}V\left(u,r\right)}{\left(V\left(u,r\right)\right)^{2}}\right)P\left(u,r\right)+\nonumber \\
 &  & +\left(-\frac{1}{4}\,\frac{\frac{\partial}{\partial r}Q\left(u,r\right)}{\pi\,r^{4}}+\frac{1}{4}\,\frac{\frac{\partial}{\partial u}Q\left(u,r\right)}{V\left(u,r\right)\pi\,r^{3}}\right)Q\left(u,r\right)\cdots
\end{eqnarray}
}
\begin{lyxcode}
\textcolor{red}{>TOVn{[}2{]}~:=~hcollect(simplify(r{*}lhs(grcomponent(TOV(dn),~{[}u{]}))/V(u,~r)),~\{P(u,~r),~rho(u,~r),~}

\textcolor{red}{diff(P(u,~r),~r),~diff(Q(u,~r),~u),~diff(V(u,~r),~r),~diff(V(u,~r),~u),~diff(rho(u,~r),~r),~}

\textcolor{red}{diff(rho(u,~r),~u),~diff(beta(u,~r),~r),~diff(beta(u,~r),~u),~diff(epsilon(u,~r),~r),~}

\textcolor{red}{diff(omega(u,~r),~r),~epsilon(u,~r),~omega(u,~r)\},~\{P(u,~r),~Q(u,~r),~epsilon(u,~r),~}

\textcolor{red}{omega(u,~r)\});}

\end{lyxcode}
\textcolor{blue}{
\begin{eqnarray}
\mathit{TOVn_{2}} & = & \left(\frac{\left(\frac{\partial}{\partial r}V\left(u,r\right)\right)\omega\left(u,r\right)}{V\left(u,r\right)\left(-1+\omega\left(u,r\right)\right)^{2}}-\frac{1}{2}\,\frac{r\frac{\partial}{\partial u}V\left(u,r\right)}{\left(V\left(u,r\right)\right)^{2}}+2\,\frac{\left(\frac{\partial}{\partial r}\beta\left(u,r\right)\right)\omega\left(u,r\right)}{\left(-1+\omega\left(u,r\right)\right)^{2}}+\frac{r\frac{\partial}{\partial u}\beta\left(u,r\right)}{V\left(u,r\right)}\right.\nonumber \\
 &  & \left.-\frac{\left(1+\omega\left(u,r\right)\right)\frac{\partial}{\partial r}\omega\left(u,r\right)}{\left(-1+\omega\left(u,r\right)\right)^{3}}+\frac{\omega\left(u,r\right)}{r\left(-1+\omega\left(u,r\right)\right)^{2}}\right)P\left(u,r\right)+\left(\frac{\left(\frac{\partial}{\partial r}V\left(u,r\right)\right)\omega\left(u,r\right)}{V\left(u,r\right)\left(-1+\omega\left(u,r\right)\right)^{2}}-\frac{1}{2}\,\frac{r\frac{\partial}{\partial u}V\left(u,r\right)}{\left(V\left(u,r\right)\right)^{2}}\right.\nonumber \\
 &  & \left.+2\,\frac{\left(\frac{\partial}{\partial r}\beta\left(u,r\right)\right)\omega\left(u,r\right)}{\left(-1+\omega\left(u,r\right)\right)^{2}}+\frac{r\frac{\partial}{\partial u}\beta\left(u,r\right)}{V\left(u,r\right)}-\frac{\left(1+\omega\left(u,r\right)\right)\frac{\partial}{\partial r}\omega\left(u,r\right)}{\left(-1+\omega\left(u,r\right)\right)^{3}}+\frac{\omega\left(u,r\right)}{r\left(-1+\omega\left(u,r\right)\right)^{2}}\right)\rho\left(u,r\right)+\nonumber \\
 &  & +\frac{\left(\frac{\partial}{\partial r}P\left(u,r\right)\right)\omega\left(u,r\right)}{\left(-1+\omega\left(u,r\right)\right)^{2}}+\frac{1}{2}\,\frac{Q\left(u,r\right)\frac{\partial}{\partial u}Q\left(u,r\right)}{r^{3}\pi\,V\left(u,r\right)}+\frac{\left(\frac{\partial}{\partial r}V\left(u,r\right)\right)\epsilon\left(u,r\right)}{V\left(u,r\right)}+\frac{\omega\left(u,r\right)\frac{\partial}{\partial r}\rho\left(u,r\right)}{\left(-1+\omega\left(u,r\right)\right)^{2}}+\frac{r\frac{\partial}{\partial u}\rho\left(u,r\right)}{V\left(u,r\right)}\nonumber \\
 &  & +2\,\left(\frac{\partial}{\partial r}\beta\left(u,r\right)\right){\color{blue}+}\epsilon\left(u,r\right)+\frac{\partial}{\partial r}\epsilon\left(u,r\right)
\end{eqnarray}
}
\begin{lyxcode}
\textcolor{red}{>TOVn{[}3{]}~:=~hcollect(simplify(lhs(grcomponent(TOV(dn),~{[}r{]}))),~\{P(u,~r),~rho(u,~r),~}

\textcolor{red}{diff(P(u,~r),~r),~diff(Q(u,~r),~r),~P{[}t{]}(u,~r)\},~\{diff(V(u,~r),~r),~diff(V(u,~r),~u),}

\textcolor{red}{diff(beta(u,~r),~r),~diff(beta(u,~r),~u)\});}\textcolor{blue}{
\begin{eqnarray}
\mathit{TOVn_{3}} & = & \left(-\frac{1}{2}\,\frac{\frac{\partial}{\partial r}V\left(u,r\right)}{V\left(u,r\right)}-\frac{r\frac{\partial}{\partial u}V\left(u,r\right)}{\left(V\left(u,r\right)\right)^{2}}-\frac{\partial}{\partial r}\beta\left(u,r\right)+2\,\frac{r\frac{\partial}{\partial u}\beta\left(u,r\right)}{V\left(u,r\right)}-\,\frac{3}{2r}\right)P\left(u,r\right)+\left(-\frac{1}{2}\,\frac{\frac{\partial}{\partial r}V\left(u,r\right)}{V\left(u,r\right)}\right.\nonumber \\
 &  & \left.-\frac{r\frac{\partial}{\partial u}V\left(u,r\right)}{\left(V\left(u,r\right)\right)^{2}}-\frac{\partial}{\partial r}\beta\left(u,r\right)+2\,\frac{r\frac{\partial}{\partial u}\beta\left(u,r\right)}{V\left(u,r\right)}+\,\frac{1}{2r}\right)\rho\left(u,r\right)-\frac{\partial}{\partial r}P\left(u,r\right)+\frac{1}{4}\,\frac{Q\left(u,r\right)\frac{\partial}{\partial r}Q\left(u,r\right)}{\pi\,r^{4}}\nonumber \\
 &  & +\frac{2P_{t}\left(u,r\right)}{r}+\frac{r\left(\frac{\partial}{\partial u}P\left(u,r\right)+\frac{\partial}{\partial u}\rho\left(u,r\right)\right)}{V\left(u,r\right)}
\end{eqnarray}
}
\end{lyxcode}
By comparing term by term, a certain regularity can be observed, and
we can verify that the two equations are the same:
\begin{lyxcode}
\textcolor{red}{>~temp03~:=~collect(simplify(TOVn{[}0{]}-TOVn{[}3{]}),~r);~}
\end{lyxcode}
\textcolor{blue}{
\begin{equation}
\mathit{temp03:=0}
\end{equation}
}With the previous expression we can conclude that
\[
e^{2\beta}T_{\;;a}^{a0}=T_{\;1;a}^{a}=0
\]
and the same conservation equation is obtained. We have in this way
3 independent conservation equations. 
\begin{lyxcode}
\textcolor{red}{>~temp21~:=~hcollect(hcollect(hcollect(simplify(TOVn{[}2{]}-TOVn{[}1{]}),~\{V(u,~r),~beta(u,~r)\},~}

\textcolor{red}{\{diff(P(u,~r),~u),~diff(V(u,~r),~r)\}),~\{P(u,~r),~rho(u,~r),~diff(P(u,~r),~r),~diff(Q(u,~r),~u)\},~}

\textcolor{red}{\{diff(rho(u,~r),~u),~diff(beta(u,~r),~r),~diff(beta(u,~r),~u)\}),~\{P(u,~r),~rho(u,~r),~}

\textcolor{red}{P{[}t{]}(u,~r)\},~\{Q(u,~r),~diff(Q(u,~r),~r)\});~}
\end{lyxcode}
\textcolor{blue}{
\begin{eqnarray}
temp21 & := & \left(-\frac{1}{2}\,\frac{\frac{\partial}{\partial r}V\left(u,r\right)}{V\left(u,r\right)}-\frac{r\frac{\partial}{\partial u}V\left(u,r\right)}{\left(V\left(u,r\right)\right)^{2}}-\frac{\partial}{\partial r}\beta\left(u,r\right)+2\,\frac{r\frac{\partial}{\partial u}\beta\left(u,r\right)}{V\left(u,r\right)}-\,\frac{3}{2r}\right)P\left(u,r\right)+\left(-\frac{1}{2}\,\frac{\frac{\partial}{\partial r}V\left(u,r\right)}{V\left(u,r\right)}\right.\nonumber \\
 &  & \left.-\frac{r\frac{\partial}{\partial u}V\left(u,r\right)}{\left(V\left(u,r\right)\right)^{2}}-\frac{\partial}{\partial r}\beta\left(u,r\right)+2{\color{blue}}\,\frac{r\frac{\partial}{\partial u}\beta\left(u,r\right)}{V\left(u,r\right)}+\,\frac{1}{2r}\right)\rho\left(u,r\right)-\frac{\partial}{\partial r}P\left(u,r\right)+\frac{1}{4}\,\frac{Q\left(u,r\right)\frac{\partial}{\partial r}Q\left(u,r\right)}{\pi\,r^{4}}\nonumber \\
 &  & +\frac{2P_{t}\left(u,r\right)}{r}+\frac{r\left(\frac{\partial}{\partial u}P\left(u,r\right)+\frac{\partial}{\partial u}\rho\left(u,r\right)\right)}{V\left(u,r\right)}
\end{eqnarray}
}
\begin{lyxcode}
\textcolor{red}{>~temp21m0~:=~collect(simplify(temp21-TOVn{[}0{]}),~r);~}
\end{lyxcode}
\textcolor{blue}{
\begin{equation}
\mathit{temp21m0:=0}
\end{equation}
}The difference of the two equations brings us back to the conservation
equation of (TOV: Tolman Oppenheimer Volkoff). However, taken separately
they are not proprocionales to $TOV[0]=TOV[3]$. 

\subsection{Einstein field equations}

Taking as input the covariant components of the Einstein field equations:
\begin{lyxcode}
\textcolor{red}{>grdef(`Eins\{a~b\}~:=~G\{a~b\}~=~8{*}Pi{*}T\{a~b\}`);~}

\textcolor{red}{>grcalc(Eins(dn,~dn));~}

\textcolor{red}{>~gralter(Eins(dn,~dn),~simplify,~factor,~radsimp);~}

\textcolor{red}{>~grdisplay(Eins(dn,~dn));~}
\end{lyxcode}
\textcolor{blue}{
\[
\begin{array}[t]{c}
\mathit{For\,the\,bondi\,spacetime:}\\
\mathit{Eins(dn,dn)}\\
\mathit{Eins(dn,dn)}
\end{array}
\]
\begin{eqnarray}
\mathit{Eins_{uu}} & = & \left(\frac{r\frac{\partial}{\partial u}V\left(u,r\right)+2\,\left(V\left(u,r\right)\right)^{2}\frac{\partial}{\partial r}\beta\left(u,r\right)-2\,rV\left(u,r\right)\frac{\partial}{\partial u}\beta\left(u,r\right)-V\left(u,r\right)\frac{\partial}{\partial r}V\left(u,r\right)+V\left(u,r\right)e^{2\,\beta\left(u,r\right)}}{r^{3}}=\right.\nonumber \\
 & = & \frac{1}{\left(-1+\omega\left(u,r\right)\right)^{2}r^{5}}\left(V\left(u,r\right){\rm e}^{2\,\beta\left(u,r\right)}\left(8\,r^{4}\pi\,\epsilon\left(u,r\right)\left(\omega\left(u,r\right)\right)^{2}+8\,r^{4}\pi\,\left(\omega\left(u,r\right)\right)^{2}\rho\left(u,r\right)+\right.\right.\nonumber \\
 &  & +8\,r^{4}\pi\,\omega\left(u,r\right)P\left(u,r\right)-16\,r^{4}\pi\,\epsilon\left(u,r\right)\omega\left(u,r\right)-8\,r^{4}\pi\,\omega\left(u,r\right)\rho\left(u,r\right)+8\,r^{4}\pi\,\epsilon\left(u,r\right)+\nonumber \\
 &  & \left.\left.+8\,r^{4}\pi\,\rho\left(u,r\right)+\left(Q\left(u,r\right)\right)^{2}-2\,\left(Q\left(u,r\right)\right)^{2}\omega\left(u,r\right)+\left(Q\left(u,r\right)\right)^{2}\left(\omega\left(u,r\right)\right)^{2}\right)\right)\nonumber \\
\mathit{Eins_{ru}} & = & \left(\frac{2\,V\left(u,r\right)\frac{\partial}{\partial r}\beta\left(u,r\right)-\frac{\partial}{\partial r}V\left(u,r\right)+e^{2\,\beta\left(u,r\right)}}{r^{2}}=\frac{e^{2\,\beta\left(u,r\right)}\left(8\,r^{4}\pi\,\rho\left(u,r\right)+\left(Q\left(u,r\right)\right)^{2}\right)}{r^{4}}\right)\nonumber \\
\mathit{Eins_{ur}} & = & \left(\frac{2\,V\left(u,r\right)\frac{\partial}{\partial r}\beta\left(u,r\right)-\frac{\partial}{\partial r}V\left(u,r\right)+e^{2\,\beta\left(u,r\right)}}{r^{2}}=\frac{e^{2\,\beta\left(u,r\right)}\left(8\,r^{4}\pi\,\rho\left(u,r\right)+\left(Q\left(u,r\right)\right)^{2}\right)}{r^{4}}\right)\nonumber \\
\mathit{Eins_{rr}} & = & \left(4\,\frac{\frac{\partial}{\partial r}\beta\left(u,r\right)}{r}=8\,\frac{\pi\,r{\rm e}^{2\,\beta\left(u,r\right)}\left(P\left(u,r\right)+\rho\left(u,r\right)\right)}{V\left(u,r\right)}\right)\nonumber \\
\mathit{Eins_{\theta\theta}} & = & \left(\frac{1}{2}\,\left(-4\,\left(\frac{\partial^{2}}{\partial u\partial r}\beta\left(u,r\right)\right)r^{2}+\left(\frac{\partial^{2}}{\partial r^{2}}V\left(u,r\right)\right)r+2\,\left(\frac{\partial}{\partial r}V\left(u,r\right)\right)\left(\frac{\partial}{\partial r}\beta\left(u,r\right)\right)r+\right.\right.\nonumber \\
 &  & \left.\left.+2\,V\left(u,r\right)\left(\frac{\partial^{2}}{\partial r^{2}}\beta\left(u,r\right)\right)r-2V\left(u,r\right)\frac{\partial}{\partial r}\beta\left(u,r\right)\right)e^{-2\,\beta\left(u,r\right)}=\frac{8\,r^{4}P_{t}\left(u,r\right)\pi+\left(Q\left(u,r\right)\right)^{2}}{r^{2}}\right)\nonumber \\
\mathit{Eins_{\phi\phi}} & = & \left(\frac{\sin\left(\theta\right)^{2}}{2}\,\left(-4\,\left(\frac{\partial^{2}}{\partial u\partial r}\beta\left(u,r\right)\right)r^{2}+\left(\frac{\partial^{2}}{\partial r^{2}}V\left(u,r\right)\right)r+2\,\left(\frac{\partial}{\partial r}V\left(u,r\right)\right)\left(\frac{\partial}{\partial r}\beta\left(u,r\right)\right)r+\right.\right.\nonumber \\
 &  & \left.+2\,V\left(u,r\right)\left(\frac{\partial^{2}}{\partial r^{2}}\beta\left(u,r\right)\right)r-2V\left(u,r\right)\frac{\partial}{\partial r}\beta\left(u,r\right)\right)e^{-2\,\beta\left(u,r\right)}=\nonumber \\
 &  & \left.\qquad\frac{\sin\left(\theta\right)^{2}\left(8\,r^{4}P_{t}\left(u,r\right)\pi+\left(Q\left(u,r\right)\right)^{2}\right)}{r^{2}}\right)
\end{eqnarray}
}

Using the simplification routines for the $uu$ component of the Einstein
tensor :$G_{00}=8\pi T_{00}$ 
\begin{lyxcode}
\textcolor{red}{>lhsEuu~:=~hcollect(lhs(grcomponent(Eins(dn,~dn),~{[}u,~u{]})),~\{r,~diff(V(u,~r),~u),~}

\textcolor{red}{diff(beta(u,~r),~u)\},~\{V(u,~r),~diff(V(u,~r),~r),~diff(beta(u,~r),~r)\})}
\end{lyxcode}
\textcolor{blue}{
\begin{eqnarray}
\mathit{lhsEuu} & = & \frac{-2\,V\left(u,r\right)\frac{\partial}{\partial u}\beta\left(u,r\right)+\frac{\partial}{\partial u}V\left(u,r\right)}{r^{2}}+\nonumber \\
 &  & \qquad+\frac{V\left(u,r\right)e^{2\,\beta\left(u,r\right)}+2\,\left(V\left(u,r\right)\right)^{2}\frac{\partial}{\partial r}\beta\left(u,r\right)-V\left(u,r\right)\frac{\partial}{\partial r}V\left(u,r\right)}{r^{3}}
\end{eqnarray}
}

And for the right side of the same component of the Einstein field
equation: 
\begin{lyxcode}
\textcolor{red}{>~rhsEuu~:=~kfactor(hcollect(hcollect(rhs(grcomponent(Eins(dn,~dn),~{[}u,~u{]})),~}

\textcolor{red}{\{Q(u,~r),~V(u,~r),~exp(2{*}beta(u,~r)),~epsilon(u,~r)\},~\{P(u,~r),~Q(u,~r),~rho(u,~r),~}

\textcolor{red}{epsilon(u,~r)\}),~\{P(u,~r),~Q(u,~r),~rho(u,~r),~epsilon(u,~r)\},~\{V(u,~r),~beta(u,~r)\}),~}

\textcolor{red}{8{*}Pi{*}exp(2{*}beta(u,~r)){*}V(u,~r)/r);~}
\end{lyxcode}
\textcolor{blue}{
\begin{eqnarray}
\mathit{rhsEuu} & := & \frac{8\,\pi\,e^{2\,\beta\left(u,r\right)}V\left(u,r\right)\left(\frac{\omega\left(u,r\right)P\left(u,r\right)}{\left(-1+\omega\left(u,r\right)\right)^{2}}+\frac{1}{8}\,\frac{\left(Q\left(u,r\right)\right)^{2}}{r^{4}\pi}+\frac{\left(\left(\omega\left(u,r\right)\right)^{2}-\omega\left(u,r\right)+1\right)\rho\left(u,r\right)}{\left(-1+\omega\left(u,r\right)\right)^{2}}+\epsilon\left(u,r\right)\right)}{r}
\end{eqnarray}
}

Let's see now its Mixed components:
\begin{lyxcode}
\textcolor{red}{>~grcalc(Eins(up,~dn));}

\textcolor{red}{>~gralter(Eins(up,~dn),~expand,~factor);~}

\textcolor{red}{>~grdisplay(Eins(up,~dn));~}
\end{lyxcode}
\textcolor{blue}{
\[
\begin{array}[t]{c}
\mathit{For\,the\,bondi\,spacetime:}\\
\mathit{Eins(up,dn)}\\
\mathit{Eins(up,dn)}
\end{array}
\]
}

\textcolor{blue}{
\begin{eqnarray*}
\mathit{Eins_{\:u}^{u}} & = & \left(\frac{2\,V\left(u,r\right)\frac{\partial}{\partial r}\beta\left(u,r\right)-\frac{\partial}{\partial r}V\left(u,r\right)+\left(e^{\beta\left(u,r\right)}\right)^{2}}{\left(e^{\beta\left(u,r\right)}\right)^{2}r^{2}}=\frac{8\,r^{4}\pi\,\rho\left(u,r\right)+\left(Q\left(u,r\right)\right)^{2}}{r^{4}}\right)\\
\mathit{Eins_{\:r}^{u}} & = & \left(4\,\frac{\frac{\partial}{\partial r}\beta\left(u,r\right)}{\left(e^{\beta\left(u,r\right)}\right)^{2}r}=8\,\frac{\pi\,r\left(P\left(u,r\right)+\rho\left(u,r\right)\right)}{V\left(u,r\right)}\right)\\
\mathit{Eins_{\:u}^{r}} & = & \left(-\frac{2\,V\left(u,r\right)\frac{\partial}{\partial u}\beta\left(u,r\right)-\frac{\partial}{\partial u}V\left(u,r\right)}{\left(e^{\beta\left(u,r\right)}\right)^{2}r^{2}}=\right.\\
 &  & \left.8\,\frac{\pi\,V\left(u,r\right)\left(\omega\left(u,r\right)P\left(u,r\right)+\epsilon\left(u,r\right)\left(\omega\left(u,r\right)\right)^{2}-2\,\epsilon\left(u,r\right)\omega\left(u,r\right)+\omega\left(u,r\right)\rho\left(u,r\right)+\epsilon\left(u,r\right)\right)}{r\left(-1+\omega\left(u,r\right)\right)^{2}}\right)\\
\mathit{Eins_{\:\theta}^{\theta}} & = & \left(\frac{1}{2}\,\frac{1}{\left(e^{\beta\left(u,r\right)}\right)^{2}r^{2}}\left(4\,\left(\frac{\partial^{2}}{\partial u\partial r}\beta\left(u,r\right)\right)r^{2}-\left(\frac{\partial^{2}}{\partial r^{2}}V\left(u,r\right)\right)r-2\,\left(\frac{\partial}{\partial r}V\left(u,r\right)\right)\left(\frac{\partial}{\partial r}\beta\left(u,r\right)\right)r+\right.\right.\\
 &  & \left.\left.-2\,V\left(u,r\right)\left(\frac{\partial^{2}}{\partial r^{2}}\beta\left(u,r\right)\right)r+2\,V\left(u,r\right)\left(\frac{\partial}{\partial r}\beta\left(u,r\right)\right)\right)=-\frac{8\,r^{4}P_{t}\left(u,r\right)\pi+\left(Q\left(u,r\right)\right)^{2}}{r^{4}}\right)\\
\mathit{Eins_{\:\phi}^{\phi}} & = & \left(\frac{1}{2}\,\frac{1}{\left(e^{\beta\left(u,r\right)}\right)^{2}r^{2}}\left(4\,\left(\frac{\partial^{2}}{\partial u\partial r}\beta\left(u,r\right)\right)r^{2}-\left(\frac{\partial^{2}}{\partial r^{2}}V\left(u,r\right)\right)r-2\,\left(\frac{\partial}{\partial r}V\left(u,r\right)\right)\left(\frac{\partial}{\partial r}\beta\left(u,r\right)\right)r+\right.\right.\\
 &  & \left.\left.-2\,V\left(u,r\right)\left(\frac{\partial^{2}}{\partial r^{2}}\beta\left(u,r\right)\right)r+2\,V\left(u,r\right)\left(\frac{\partial}{\partial r}\beta\left(u,r\right)\right)\right)=-\frac{8\,r^{4}P_{t}\left(u,r\right)\pi+\left(Q\left(u,r\right)\right)^{2}}{r^{4}}\right)
\end{eqnarray*}
}Grouping terms for the right side of the EqnEins $\mathit{Eins(up,dn)[r,u]}$
component has 
\begin{lyxcode}
\textcolor{red}{>~rhsEru~:=~kfactor(hcollect(rhs(grcomponent(Eins(up,~dn),~{[}r,~u{]})),~\{P(u,~r),~}

\textcolor{red}{epsilon(u,~r),omega(u,~r)\},~\{V(u,~r)\}),~8{*}V(u,~r){*}Pi/r);~}
\end{lyxcode}
\textcolor{blue}{
\begin{eqnarray}
\mathit{rhsEru} & = & \frac{8\,\pi\,V\left(u,r\right)}{r}\left(\frac{\omega\left(u,r\right)P\left(u,r\right)}{\left(-1+\omega\left(u,r\right)\right)^{2}}+\epsilon\left(u,r\right)+\frac{\omega\left(u,r\right)\rho\left(u,r\right)}{\left(-1+\omega\left(u,r\right)\right)^{2}}\right)
\end{eqnarray}
}An interesting expression is the one associated with the field equation
$\left(G_{11}=8\pi T_{11}\right)$, which includes the pressure and
is very similar to the one obtained previously $\left(G_{\,1}^{1}=8\pi T_{\,1}^{1}\right)$,
which relates density to the functions of the metric 
\begin{lyxcode}
\textcolor{red}{>~lhsErr~:=~hcollect(lhs(grcomponent(Eins(up,~dn),~{[}r,~r{]})),~\{P(u,~r),~}

\textcolor{red}{epsilon(u,~r),~omega(u,~r)\},~\{V(u,~r)\});~}
\end{lyxcode}
\textcolor{blue}{
\begin{equation}
\mathit{lhsErr}:=\frac{-2\,V\left(u,r\right)\frac{\partial}{\partial r}\beta\left(u,r\right)-\frac{\partial}{\partial r}V\left(u,r\right)+\left(e^{\beta\left(u,r\right)}\right)^{2}}{\left(e^{\beta\left(u,r\right)}\right)^{2}r^{2}}
\end{equation}
}
\begin{lyxcode}
\textcolor{red}{>~rhsErr~:=~hcollect(rhs(grcomponent(Eins(up,~dn),~{[}r,~r{]})),~\{P(u,~r)\},~\{V(u,~r)\});}
\end{lyxcode}
\textcolor{blue}{
\begin{equation}
\mathit{rhsErr}:=-8\,\pi\,P\left(u,r\right)+\frac{\left(Q\left(u,r\right)\right)^{2}}{r^{4}}
\end{equation}
}Both relate the metrical elements -{}- $V(u,r)\,and\,\beta\left(u,r\right)$
-{}- with the physical quantities -{}- $P(u,r)\,and\,\rho(u,r)$-{}-
of pressure and density.

\subsection{Trace}

Let's calculate the trace of G
\begin{lyxcode}
\textcolor{red}{>~grdef(`TG~:=~g\{\textasciicircum a~\textasciicircum b\}{*}G\{a~b\}~`);}

\textcolor{red}{>~grcalcd(TG);}\textcolor{blue}{
\[
\begin{array}[t]{c}
\mathit{`CPU\,Time`=0.005}\\
\mathit{For\,the\,bondi\,spacetime:}\\
\mathit{TG}
\end{array}
\]
}
\end{lyxcode}
\textcolor{blue}{
\begin{eqnarray}
\mathit{TG} & := & \frac{1}{e^{2\,\beta\left(u,r\right)}r^{2}}\left(2\,V\left(u,r\right)\frac{\partial}{\partial r}\beta\left(u,r\right)-2\,\frac{\partial}{\partial r}V\left(u,r\right)+2\,e^{2\,\beta\left(u,r\right)}+4\,\left(\frac{\partial^{2}}{\partial u\partial r}\beta\left(u,r\right)\right)r^{2}\right.\nonumber \\
 &  & \left.-\left(\frac{\partial^{2}}{\partial r^{2}}V\left(u,r\right)\right)r-2\,\left(\frac{\partial}{\partial r}V\left(u,r\right)\right)\left(\frac{\partial}{\partial r}\beta\left(u,r\right)\right)r-2\,V\left(u,r\right)\left(\frac{\partial^{2}}{\partial r^{2}}\beta\left(u,r\right)\right)r\right)
\end{eqnarray}
}Very similar to the equation the left side of the angular component,
of the field equations $G_{2}^{2}=8\pi T_{2}^{2}$. Now the trace
of the Stress tensor:
\begin{lyxcode}
\textcolor{red}{>~grdef(`TT~:=~g\{\textasciicircum a~\textasciicircum b\}{*}T\{a~b\}~`);~}

\textcolor{red}{>~grcalcd(TT);~}\textcolor{blue}{
\[
\begin{array}[t]{c}
\mathit{`CPU\,Time`=0.004}\\
\mathit{For\,the\,bondi\,spacetime:}\\
\mathit{TT}
\end{array}
\]
\begin{equation}
TT=\rho\left(u,r\right)-P\left(u,r\right)-2P_{t}\left(u,r\right)
\end{equation}
}
\end{lyxcode}
The contraction of the Field equation $\mathit{Traza=g^{\mu\nu}G_{\mu\nu}=8\pi g^{\mu\nu}T_{\mu\nu}}$
\begin{lyxcode}
\textcolor{red}{>~grdef(`TrazaTE~:=~g\{\textasciicircum a~\textasciicircum b\}{*}G\{a~b\}~=~8{*}Pi{*}~g\{\textasciicircum a~\textasciicircum b\}{*}T\{a~b\}`);~}

\textcolor{red}{>~grcalcd(TrazaTE);}\textcolor{blue}{
\begin{eqnarray}
\mathit{TrazaTE} & = & \frac{1}{e^{2\,\beta\left(u,r\right)}r^{2}}\left(2\,V\left(u,r\right)\frac{\partial}{\partial r}\beta\left(u,r\right)-2\,\frac{\partial}{\partial r}V\left(u,r\right)+2\,e^{2\,\beta\left(u,r\right)}+4\,\left(\frac{\partial^{2}}{\partial u\partial r}\beta\left(u,r\right)\right)r^{2}\right.\nonumber \\
 &  & \left.-\left(\frac{\partial^{2}}{\partial r^{2}}V\left(u,r\right)\right)r-2\,\left(\frac{\partial}{\partial r}V\left(u,r\right)\right)\left(\frac{\partial}{\partial r}\beta\left(u,r\right)\right)r-2\,V\left(u,r\right)\left(\frac{\partial^{2}}{\partial r^{2}}\beta\left(u,r\right)\right)r\right)\nonumber \\
 & = & -8\,\pi\,\left(-\rho\left(u,r\right)+P\left(u,r\right)+2\,P_{t}\left(u,r\right)\right)
\end{eqnarray}
}
\end{lyxcode}
This equation of the trace is important since it can be shown that
it is equivalent to the equation $\left(T_{\quad;\mu}^{\mu0}=0\right)$,
the Tolman-Oppenheimer-Volkoff equation. When comparing the left side
of the trace equation with the left side of the angular component
$G_{\,2}^{2}=8\pi T_{\:2}^{2}$, of the equations from field, we get
\begin{lyxcode}
\textcolor{red}{>difG22TG~:=~simplify((1/2){*}grcomponent(TG)-lhs(grcomponent(Eins(up,~dn),~{[}theta,~theta{]})))}
\end{lyxcode}
\textcolor{blue}{
\begin{equation}
difG22TG:=\frac{e^{-2\,\beta\left(u,r\right)}\left(-\frac{\partial}{\partial r}V\left(u,r\right)+e^{2\,\beta\left(u,r\right)}\right)}{r^{2}}
\end{equation}
}

\section{\label{sec:Results}Results and Comments }

This section is dedicated to presenting some results obtained when
executing the established and described commands \emph{(MAPLE17 +
GRTensorIII)} of the previous section. We can summarize that the spreadsheet
allows us to calculate the different tensor terms for a radiative
coordinate system from the Einstein Field equations and their Conservation
equations. Throughout the spreadsheet development process, the importance
of the expression to determine the components of the Stress tensor
from the Minkowski metric to its connection with the Bondi radiative
coordinate metric of the subsection (\ref{subsec:BRiCS}) should be
taken into account (\ref{eq:Mink2Bondi}). You can see that it is
necessary to specify the scope of each component of the metric (\ref{eq:Mink2Bondi}),
to make the calculation of the entire sheet flow without interruption
(\citep{1996CQGra..13.1885M}). This procedure has been used to perform
the calculations for (\citep{2018PPhy..14.1..46M,2022InJPh..96..317M}).
We prove that this calculation process fully coincides with the result
obtained manually. In these articles the equations allow us to carry
out the study and behavior of the physical variables calculated in
the articles cited above. For example, the Conservation equations$\left(T_{\nu;\mu}^{\mu}=0\right)$
for the charged case are:

\begin{eqnarray}
T_{0;\lambda}^{\lambda} & = & \frac{V}{r^{2}}\left(1+\frac{rV_{,1}}{V}\right)\left[\varepsilon+\frac{\omega\left(\rho+P\right)}{\left(1-\omega\right)^{2}}\right]+\nonumber \\
 &  & +\frac{V}{r}\frac{\partial}{\partial r}\left[\varepsilon+\frac{\omega\left(\rho+P\right)}{\left(1-\omega\right)^{2}}\right]+\frac{\partial}{\partial u}\left(\rho+\frac{Q^{2}}{8\pi r^{4}}\right)=0\\
T_{1;\lambda}^{\lambda} & = & \frac{e^{-2\beta}}{2\pi r}\beta_{,10}-\frac{\partial}{\partial r}\left(P-\frac{Q^{2}}{8\pi r^{4}}\right)-\frac{1}{2}\left(2\beta_{,1}+\frac{V_{,1}}{V}-\frac{1}{r}\right)\left(\rho+P\right)+\nonumber \\
 &  & -\frac{2}{r}\left[\left(P-\frac{Q^{2}}{8\pi r^{4}}\right)-\left(P_{t}+\frac{Q^{2}}{8\pi r^{4}}\right)\right]=0\\
T_{\,;\lambda}^{\lambda0} & = & e^{-2\beta}T_{1;\lambda}^{\lambda}=0\\
e^{2\beta}T_{\,;\lambda}^{\lambda1} & = & \frac{V}{r^{2}}\left(1+4r\beta_{,1}+\frac{rV_{,1}}{V}\right)\left[\varepsilon+\frac{\omega\left(\rho+P\right)}{\left(1-\omega\right)^{2}}\right]+\frac{V}{2r}\left(2\beta_{,1}+\frac{V_{,1}}{V}-\frac{1}{r}\right)\left(\rho+P\right)+\nonumber \\
 &  & +\frac{2V}{r^{2}}\left[\left(P-\frac{Q^{2}}{8\pi r^{4}}\right)-\left(P_{t}+\frac{Q^{2}}{8\pi r^{4}}\right)\right]+\nonumber \\
 &  & \frac{V}{r}\frac{\partial}{\partial r}\left(P-\frac{Q^{2}}{8\pi r^{4}}\right)+\frac{V}{r}\frac{\partial}{\partial r}\left[\varepsilon+\frac{\omega\left(\rho+P\right)}{\left(1-\omega\right)^{2}}\right]-\frac{\partial}{\partial u}\left(P-\frac{Q^{2}}{8\pi r^{4}}\right)=0
\end{eqnarray}
As can be seen in (\ref{subsec:(TOV)}) and the concordance of these
results with those obtained manually, it can allow us to trust that
this calculation to be carried out in another coordinate system should
have a similar performance. Of course, it is necessary to point out
that it is still pending to use the complete equation $\left(31\right)$
shown in the article (\citep{2022InJPh..96..317M}). This comparative
study between the results obtained so far and solving the complete
equation with the terms $\frac{\partial\varepsilon}{\partial r}\:and\:\frac{\partial\omega}{\partial r}$
will be carried out in a future work to publish.

\section{\label{sec:Conclusions}Conclusions }

In this article we describe a series of computer procedures used in
GR relying on the facilities of an integrated platform such as the
\emph{Maple} package and \emph{GRTensorIII.} These procedures of computational
algebra, numerical and graphic computation can facilitate algebraic
calculation both for research purposes and for teaching GR at different
levels.

\bibliographystyle{unsrtnat}

\end{document}